\documentclass[journal,twoside]{IEEEtran}

\usepackage{amsmath,amssymb,amsfonts}
\usepackage{algorithmic}
\usepackage{graphicx}
\usepackage{textcomp}
\usepackage{bm}
\usepackage[T1]{fontenc}
\usepackage{tabularx}

\usepackage{color}
\usepackage{babel}
\usepackage{float}
\usepackage{booktabs}
\usepackage{amsmath}
\usepackage{amssymb}
\usepackage{stackrel}

\usepackage[caption=false]{subfig}

\begin{document}

\title{A Dual-domain Regularization Method for Ring Artifact Removal of X-ray CT}

\author{Hongyang Zhu$^\dag$, Xin Lu$^\dag$, Yanwei Qin, Xinran Yu, Tianjiao Sun, and Yunsong Zhao$^*$.
\thanks{This work is supported in part by the National Natural Science Foundation of China (NSFC) (62271330, 61871275, 61771324, and 61827809) and National Key R\&D Program of China (2023YFA1011400).}
\thanks{The authors are with the School of Mathematical Sciences, Capital Normal University, Beijing, 100048, China, and are with Beijing Advanced Innovation Center for Imaging Technology, Capital Normal University, Beijing, 100048, China. (e-mail: 2210502076@cnu.edu.cn, luxin0808@126.com, qinyanwei2013@163.com,\ yxr6500@163.com, stj070102@163.com, zhao\_yunsong@cnu.edu.cn).}
\thanks{$^\dag$Hongyang Zhu and Xin Lu contribute equally to this work.}
\thanks{*Corresponding author: Yunsong Zhao.}}

\maketitle

\begin{abstract}
Ring artifacts in computed tomography (CT) images, arising from the undesirable responses of detector units, significantly degrade image quality and diagnostic reliability. To address this challenge, we propose a dual-domain regularization model to effectively remove ring artifacts, while maintaining the integrity of the original CT image. The proposed model corrects the vertical stripe artifacts on the sinogram by innovatively updating the response inconsistency compensation coefficients of detector units, which is achieved by employing the group sparse constraint and the projection-view direction sparse constraint on the stripe artifacts. Simultaneously, we apply the sparse constraint on the reconstructed image to further rectiﬁed ring artifacts in the image domain. The key advantage of the proposed method lies in considering the relationship between the response inconsistency compensation coefficients of the detector units and the projection views, which enables a more accurate correction of the response of the detector units. An alternating minimization method is designed to solve the model. Comparative experiments on real photon counting detector (PCD) data demonstrate that the proposed method not only surpasses existing methods in removing ring artifacts but also excels in preserving structural details and image fidelity.
\end{abstract}



\section{Introduction}
\IEEEPARstart{P}{hoton} counting detector (PCD) CT has been one of
the hot topics in X-ray imaging in recent years \cite{taguchi2013vision,willemink2018photon,yu2016evaluation,schmidt2017spectral}. Though PCD has promising potential for offering significant improvements to existing CT imaging techniques, challenges remain for the wider application of PCD CT imaging. One of the challenges in the development of PCD CT is that the location of the energy thresholds tends to vary among detector units \cite{persson2012framework}. If not compensated for, this threshold variation leads to stripe artifacts in the sinogram
and subsequent concentric ring artifacts in the reconstructed CT image. These ring artifacts can greatly degrade the quality of the CT images, hamper quantitative analysis, and reduce diagnostic accuracy. With the wide range of application requirements, there is an urgent need to develop an efficient ring artifact removal method.

Existing ring artifact removal methods can be roughly categorized
into hardware-based methods and software-based methods. Hardware-based methods mainly change the relative positions of the components of the CT system and the scanned sample and spread out the effects of response inconsistencies of detector units on the reconstructed CT images. Davis et al. \cite{davis1997x} move the detector or the sample during data acquisition, which leads to significantly reduced ring artifacts. Zhu et al. \cite{zhu2013micro} use a similar method to eliminate ring artifacts caused by the defects of the scintillator or CCD camera in optically coupled detectors. However, these methods require additional motion axes to support the movement of the ray
source, detector, or scanned sample, which often increases manufacturing costs and may introduce additional artifacts in the reconstructed CT images, such as amplified noises \cite{liu2023detector}. Consequently, additional algorithms are often required to further correct these artifacts.

Compared to hardware-based methods, software-based methods are generally easier to implement. Software-based methods can be broadly categorized into flat-field correction methods, pre-processing methods, post-processing methods, and dual-domain processing methods. The flat-field correction methods \cite{van2015dynamic,kwan2006improved,herman2009fundamentals,lifton2019ring,seibert1998flat,bangsgaard2023low} are the common methods for ring artifact correction. In these methods, an air scan is accompanied by a sample scan. The air scan incorporates the effects of inhomogeneity in the X-ray beam, e.g., the inhomogeneous response of the scintillator and the detector of the CT system. These methods are simple to implement and apply to most samples, however, they could only reduce the ring artifacts to some extent, and ring artifacts could not be completely removed since ring artifacts are influenced by various factors, such as the X-ray spectrum and the scanned sample. Pre-processing methods primarily aim to eliminate ring artifacts by removing vertical stripe artifacts on the sinogram. Various stripe artifact correction methods have been proposed, typically including filtering-based methods and mean projection-based methods. The filtering-based methods usually employ image transformations to identify the position of the vertical
stripes on the sinogram first, and then apply interpolation or filtering methods to correct the sinogram. Raven et al. \cite{raven1998numerical} remove vertical stripe artifacts by applying the Fourier transform to convert the sinogram into the frequency domain, and then filter the high-frequency components associated with the vertical stripe artifacts through the application of low-pass filtering. Münch et al. \cite{munch2009stripe} proposed an improved version of this method, which is based on the combination of wavelet and Fourier filtering (WF). The WF method performs Fourier filtering on the coefficients of 2-D wavelet decomposed vertical detail band instead of the original sinogram. It utilizes the multi-scale properties of the wavelet transform to more effectively identify vertical stripes, especially those with larger widths. However, these methods achieve ring artifact removal by suppressing specific frequency components in the frequency domain, though effective in eliminating artifacts, may inadvertently suppress
essential frequency components of the sinogram, leading to potential loss of CT image detail. The mean projection-based methods first employ the inverse of the sinogram's mean vector, calculated along the projection view direction, to correct vertical stripe artifacts in the sinogram. Subsequently, they utilize a filtering method to compensate for any information that may have been diminished in the sinogram during the correction process. Wu et al. \cite{an2020ring} propose a $\text{TV-L}^{1}$aG
filter to filter the mean vector and then compensate the smoothed
mean vector into the sinogram. Eldib et al. \cite{eldib2017ring}
and Podgorsak et al. \cite{podgorsak2018use} implement a strategy
where they filter the difference images of the reconstructed image
and the sinogram before and after the correction, respectively. The filtered difference image is then recombined with the corrected reconstructed image or sinogram to compensate for any lost information. However, the effectiveness of these methods relies on the accurate localization and compensation of vertical stripe artifacts. Incomplete correction or inaccurate compensation may lead to the loss of important information.

As for post-processing methods, since ring artifacts in Cartesian
coordinates manifest as vertical stripe artifacts in polar coordinates which are generally easier to identify, most of the methods are developed based on polar coordinate transformations. These methods commence by converting the reconstructed image from Cartesian coordinates to polar coordinates, a process during which the concentric ring artifacts are converted into vertical stripe artifacts. Subsequently, image processing methods are applied to detect and remove these vertical stripe artifacts within the polar coordinate image. The final stage involves reverting the corrected image from polar coordinates back
to Cartesian coordinates. Chen et al. \cite{chen2009ring} decompose images into different independent components using independent component analysis (ICA), and only components containing stripe artifacts are selected to be filtered. Yan et al. \cite{yan2016variation} propose a variation-based ring artifact correction method with sparse constraints on the images in the polar coordinates. Liang et al. \cite{liang2017iterative}
apply the relative total variance to remove stripe artifacts in polar coordinate images, and an iterative framework is then applied to further compensate for the normal tissue details and remove the artifacts. However, a common challenge encountered by these methods is the coordinate transformation. While it is instrumental in the detection and correction of ring artifacts, it also poses a risk of degrading the spatial resolution
of the reconstructed CT images. Striking a balance between artifact removal and preservation of image resolution remains a critical challenge in the application of these post-processing methods.

Dual-domain methods are a series of methods that remove vertical stripe artifacts on the sinogram and the ring artifacts on the reconstructed CT images simultaneously. Most recently developed deep-learning-based methods belong to this category \cite{fang2019comparison,yuan2021deep,wang2019removing,fu2023deep}. These methods are usually supervised methods and rely on training datasets, posing challenges in maintaining consistency between numerically simulated data and real data. Compared to pre-processing methods and post-processing methods, dual-domain methods show better ring artifact removals ability, such as the iterative method proposed by Salehjahromi et al. \cite{salehjahromi2019new}, which aims to correct ring artifacts
by iteratively optimizing the clean reconstructed image and compensation coefficient vector. However, this method uses the same compensation coefficient vector for scan data at different projection views and may result in unsatisfactory results when dealing with vertical stripe artifacts that exhibit significant intensity variations in the projection view direction.

To address the limitations of the existing methods, we propose a novel dual-domain regularization model to effectively remove ring artifacts. The proposed model corrects the vertical stripe artifacts on the sinogram by innovatively updating the response inconsistency compensation coefficients of detector units, which is achieved by employing the group sparse constraint and the projection-view direction sparse constraint on the stripe artifacts. In addition, the ring artifacts in the image domain
are further rectified through the imposition of sparse constraint
on the reconstructed CT image. The key advantage of the proposed method lies in considering the relationship between the response inconsistency compensation coefficients of the detector units and the projection views, which enables a more accurate correction of the response of the detector units. The effectiveness of the method is evaluated through numerical experiments conducted on real data obtained by the PCD. The experimental results demonstrate that our method is capable of effectively removing ring artifacts in the reconstructed image while preserving the structure and details of the image when compared to other methods used for comparison.

\section{Materials and methods}

\subsection{Formulation for Ring Artifact Removal}

Take the fan-beam CT scan as an example to formulate the ring artifact removal framework. Let $\boldsymbol{I}_{i}\in \mathbb{R}^{V\times1}$ $(i=1,\cdots,U)$ be the scan data of the $i$th detector unit at different projection views, where $V$ be the number of projection views and $U$ be the number of detector units, and let $F_{i}$ and $D_{i}$ be the corresponding flat-field data and dark-field data, respectively; then the normalized scan data
$\boldsymbol{q}_{i}\in \mathbb{R}^{V\times1}$ can be calculated with formulae: 
\begin{equation}
\boldsymbol{q}_{i}=\frac{\boldsymbol{I}_{i}-D_{i}}{F_{i}-D_{i}}\label{eq:(1)}
\end{equation}

Because of the inconsistent response of the detector units, some of the normalized scan data $\boldsymbol{q}_{i}$ should be compensated by the response inconsistency compensation coefficients $\boldsymbol{s}_{i}\in \mathbb{R}^{V\times1}$ to obtain the corresponding ideal normalized scan data $\widetilde{\boldsymbol{q}}_{i}$, thus we have:
\begin{equation}
\widetilde{\boldsymbol{q}}_{i}=\boldsymbol{s}_{i}\odot\boldsymbol{q}_{i}\label{eq:(2)}
\end{equation}

Taking the negative logarithm on both sides of (\ref{eq:(2)}), we can obtain the clean sinogram $\boldsymbol{p}=[\boldsymbol{p}_{1},\cdots,\boldsymbol{p}_{i},\cdots,\boldsymbol{p}_{_U}]\in \mathbb{R}^{V\times U}$, where $\boldsymbol{p}_{i}\in \mathbb{R}^{V\times1}$ is defined as follows:
\begin{equation}
\boldsymbol{p}_{i}\triangleq-\log\widetilde{\boldsymbol{q}}_{i}=-\log\boldsymbol{s}_{i}-\log\boldsymbol{q}_{i}\triangleq\boldsymbol{S}_{i}+\boldsymbol{y}_{i}\label{eq:(3)}
\end{equation}
where $\boldsymbol{y}_{i}\in \mathbb{R}^{V\times1}$ is a column of the real sinogram $\boldsymbol{y}=[\boldsymbol{y}_{1},\cdots,\boldsymbol{y}_{i},\cdots,\boldsymbol{y}_{_U}]\in \mathbb{R}^{V\times U}$ and $\boldsymbol{S}_{i}\in \mathbb{R}^{V\times1}$ is the response inconsistency compensation coefficients in the projection domain. Denoting that $\boldsymbol{s}=[\boldsymbol{s}_{1},\cdots,\boldsymbol{s}_{i},\cdots,\boldsymbol{s}_{_U}]\in \mathbb{R}^{V\times U}$ and $\boldsymbol{S}=[\boldsymbol{S}_{1},\cdots,\boldsymbol{S}_{i},\cdots,\boldsymbol{S}_{_U}]\in \mathbb{R}^{V\times U}$ are compensation coefficients for all detector units. Therefore, if we can accurately estimate the compensation coefficients $\boldsymbol{s}$ or $\boldsymbol{S}$, we can obtain the clean sinogram $\boldsymbol{p}$ by (\ref{eq:(2)}) or (\ref{eq:(3)}).

\textcolor{black}{Typically, the corrected reconstructed CT image vector $\boldsymbol{x}\in \mathbb{R}^{N\times1}$ ($N$ denotes the number of image pixels) as well as the response inconsistency compensation coefficients $\boldsymbol{S}$ can be obtained by solving the following optimization model: 
\begin{equation}
\underset{\boldsymbol{x},\boldsymbol{S}}{\text{argmin}}\left\{ \begin{array}{c}
\cfrac{1}{2}\stackrel[i=1]{U}{\sum}\left\Vert A_{i}\boldsymbol{x}-\boldsymbol{p}_{i}+\boldsymbol{S}_{i}\right\Vert \\+\lambda_{1}{R}_{I}\left(\boldsymbol{x}\right)
+\lambda_{2}{R}_{P}\left(\boldsymbol{S}\right)
\end{array}\right\} \label{eq:(4)}
\end{equation}
where $A_{i}\in \mathbb{R}^{V\times N}$ represents the system matrix corresponding to the $i\text{th}$ detector unit; ${R}_{I}\left(\boldsymbol{x}\right)$ is the regularization term about the reconstructed CT image, and ${R}_{P}\left(\boldsymbol{S}\right)$ is the regularization term about the compensation coefficients; $\lambda_{1}$ and $\lambda_{2}$ are regularization parameters; The design of ${R}_{I}\left(\boldsymbol{x}\right)$ and ${R}_{P}\left(\boldsymbol{S}\right)$ plays a crucial role in the ring artifact removal effect.}

\subsection{Related methods}

\textcolor{black}{In practice, various factors would affect the correction coefficients $\boldsymbol{s}$ or $\boldsymbol{S}$, such as scanning voltage, current, filter, and the scanned sample \cite{persson2012framework}. So, the correction coefficients $\boldsymbol{s}$ or $\boldsymbol{S}$
are generally projection view-dependent. But for simplicity, many ring artifact removal methods simplify the model by assuming that all elements of $\boldsymbol{s}_{i}$ or $\boldsymbol{S}_{i}$ are
the same, i.e. $\boldsymbol{s}_{i}=s_{i}\boldsymbol{1}$, $\boldsymbol{S}_{i}=S_{i}\boldsymbol{1}$, and $\boldsymbol{1}\in \mathbb{R}^{V\times1}$ is a vector whose elements are also `1', such as the methods proposed by Wu }et al.\textcolor{black}{{}
\cite{an2020ring} and Salehjahromi }et al.\textcolor{black}{{} \cite{salehjahromi2019new}.
With this assumption, the compensation coefficients for all detector units are simplified from a matrix to a row vector $\boldsymbol{s}=[s_{1},\cdots,s_{i},\cdots,s_{_U}]\in \mathbb{R}^{1\times U}$ and $\boldsymbol{S}=[S_{1},\cdots, S_{i},\cdots, S_{_U}]\in \mathbb{R}^{1\times U}$.}

\textcolor{black}{Wu }et al.\textcolor{black}{{} \cite{an2020ring} propose a mean projection-based method to estimate the correction coefficient vector $\boldsymbol{s}=[s_{1},\cdots,s_{i},\cdots, s_{_U}]\in \mathbb{R}^{1\times U}$. Suppose that $\boldsymbol{m}\in \mathbb{R}^{1\times U}$ is the real mean projection consisting of the mean value $m_{i}$ of scanned data $\boldsymbol{q}_{i}$, and $\widetilde{\boldsymbol{m}}\in \mathbb{R}^{1\times U}$ is the ideal mean projection, which can be approximately estimated by operating $\text{TV-L}^{1}$ filtering and Gaussian filtering on $\boldsymbol{m}$. Therefore, the compensation coefficient vector $\boldsymbol{s}$ can be calculated by the following equation:
\begin{equation}
\boldsymbol{s}=\frac{\widetilde{\boldsymbol{m}}}{\boldsymbol{m}}\label{eq:(5)}
\end{equation}}

\textcolor{black}{Salehjahromi }et al.\textcolor{black}{{} \cite{salehjahromi2019new} proposes a dual-domain regularization model to estimate the compensation
coefficients $\boldsymbol{S}=[S_{1},\cdots,S_{i},\cdots, S_{_U}]\in \mathbb{R}^{1\times U}$ in the projection domain. In this model, a novel ring total variation regularization ${R}_{I}(\boldsymbol{x})=\text{RTV}_{\alpha,\theta}(\boldsymbol{x})$
is developed to correct the ring artifacts in the image domain. Moreover, to correct the sinogram, a projection domain regularization ${R}_{P}\left(\boldsymbol{S}\right)=\stackrel[i=1]{U}{\sum}\left\Vert \boldsymbol{S}\right\Vert _{1}$
is proposed to estimate the compensation coefficient vector. The ring artifacts in the reconstructed image are gradually corrected and the compensation coefficient vector is updated by applying an alternating
minimization method.}

\textcolor{black}{All of the above methods are effective in suppressing ring artifacts. However, the fact that they apply the same compensation coefficient to the scan data of the same detector unit in different
projection views, leads to the limitation in removing those vertical stripe artifacts that show significant differences in different projection view directions.}

\subsection{The proposed optimization model}

In this paper, we aim to overcome the limitations of existing methods by proposing a novel dual-domain regularization ring artifact removal model. In the proposed model, we design a specific projection domain regularization term ${R}_{P}\left(\boldsymbol{S}\right)$, which is achieved by exploiting the structural characteristics of vertical stripe artifacts present, to estimate the compensation coefficient matrix $\boldsymbol{S}\in \mathbb{R}^{V\times U}$. Moreover, we introduce an image domain regularization term ${R}_{I}\left(\boldsymbol{x}\right)$ to further correct the ring artifacts in the image domain according to the features of the reconstructed CT image. We provide an in-depth description of the proposed optimization model below.

\subsubsection{The projection domain regularization term ${R}_{P}\left(\boldsymbol{S}\right)$}

we consider incorporating the correlation between the compensation
coefficient matrix $\boldsymbol{S}\in \mathbb{R}^{V\times U}$ with the projection
views within the optimization framework. The basic assumption is that
although the response inconsistency of the detector units is related
to the projection views, the energy spectrum of X-rays received by
the same detector unit only changes slightly across a subset of adjacent
views. Therefore, the response inconsistency compensation coefficients
of the detector units can be assumed to be constant in a subset of
adjacent views. Hence, the gradient of the compensation coefficients
along the projection view direction is sparse, leading to the first
regularization term in the projection domain to provide a sparse solution
for $\boldsymbol{S}$ along the projection views:

\begin{equation}
R_{P1}(\boldsymbol{S})=\left\Vert \nabla_{V}\boldsymbol{S}\right\Vert _{1}\label{eq:(6)}
\end{equation}
where $\nabla_{V}$ is the vertical difference operator defined by 
\[
\left[\nabla_{V}\boldsymbol{S}\right]_{j,i}=\begin{cases}
S_{j+1,i}-S_{j,i} & j=1,\ldots,V-1\\
S_{1,i}-S_{j,i} & j=V
\end{cases}
\]
and $\left\Vert \boldsymbol{u}\right\Vert _{1}=\stackrel[i=1]{U}{\sum}\stackrel[j=1]{V}{\sum}\left|u_{j,i}\right|$
denotes the $\text{L}_{1}$-norm of $\boldsymbol{u}$.

In addition, it is well known that the number of detector units with
non-ideal responses in practical scenarios should be limited. Therefore,
we consider \textcolor{black}{$\boldsymbol{S}_{i}$ as a group and
utilize the group sparsity constraint \cite{chen2019hyperspectral}
to provide a column sparse solution for }$\boldsymbol{S}\in \mathbb{R}^{V\times U}$,
leading to the second regularization term in the projection domain:
\begin{equation}
R_{P2}(\boldsymbol{S})=\left\Vert \boldsymbol{S}\right\Vert _{2,1}\label{eq:(7)}
\end{equation}
Here $\left\Vert \boldsymbol{u}\right\Vert _{2,1}=\stackrel[i=1]{U}{\sum}\sqrt{\stackrel[j=1]{V}{\sum}u_{j,i}^{2}}$.

\subsubsection{The image domain regularization term ${R}_{I}\left(\boldsymbol{x}\right)$}

To tackle the challenge of residual artifacts that persist when relying only on projection domain information, we incorporate supplementary constraints on the reconstructed CT image to enhance the correction of ring artifacts within the image domain. A widely adopted assumption about the ideal CT image is that it can be well approximated by a piecewise constant function. This motivates us to construct the following image domain regularization term to further rectify the ring artifacts and noise on the reconstructed CT image: 
\begin{equation}
{R}_{I}\left(\boldsymbol{x}\right)=\left\Vert \nabla_{H}\boldsymbol{x}\right\Vert _{1}+\left\Vert \nabla_{V}\boldsymbol{x}\right\Vert _{1}\label{eq:(8)}
\end{equation}
where $\nabla_{H}$ is the horizontal difference operator defined by 
\[
\left[\nabla_{H}\boldsymbol{S}\right]_{j,i}=\begin{cases}
S_{j,i+1}-S_{j,i} & i=1,\ldots,U-1\\
S_{j,1}-S_{j,i} & i=U
\end{cases}
\]

It should be pointed out that we only use the anisotropic total variation (ATV) regularization term as an example to compute the piecewise constant approximation, which can also be achieved by introducing regularization terms such as BM3D
\cite{dabov2007image}, non-local means filtering \cite{buades2005non}, and so on.

In conclusion, the ring artifact removal model proposed in this paper
can be summarized as follows: 
\begin{equation}
\underset{\boldsymbol{x},\boldsymbol{S}}{\text{argmin}}\left\{ \begin{array}{c}
\cfrac{1}{2}\stackrel[i=1]{U}{\sum}\left\Vert A_{i}\boldsymbol{x}-\boldsymbol{p}_{i}+\boldsymbol{S}_{i}\right\Vert \\
+\lambda_{1}\left(\left\Vert \nabla_{H}\boldsymbol{x}\right\Vert _{1}+\left\Vert \nabla_{V}\boldsymbol{x}\right\Vert _{1}\right)\\
+\lambda_{2}\left\Vert \nabla_{V}\boldsymbol{S}\right\Vert _{1}+\lambda_{3}\left\Vert \boldsymbol{S}\right\Vert _{2,1}
\end{array}\right\} \label{eq:(9)}
\end{equation}
where $\lambda_{1},\lambda_{2}$ and $\lambda_{3}$ are regularisation
parameters.

\subsection{Numerical implementation}

Since the optimization model (\ref{eq:(9)}) is non-convex, the solving
strategy significantly impacts the results, and only locally optimal
solutions can be expedited. In this paper, an alternating minimization
method is employed to solve (\ref{eq:(9)}).

Assuming that the estimate image $\boldsymbol{x}^{(k)}$ and $\boldsymbol{S}^{(k)}$
is obtained after $k$ iterations, then $\boldsymbol{x}^{(k+1)}$
and $\boldsymbol{S}^{(k+1)}$ can be obtained by alternately solving
the following two minimization problems:

Sub-problem 1 : 
\begin{equation}
\boldsymbol{x}^{(k+1)}=\underset{\boldsymbol{x}}{\text{argmin}}\left\{ \begin{array}{c}
\cfrac{1}{2}\stackrel[i=1]{U}{\sum}\left\Vert A_{i}\boldsymbol{x}-\boldsymbol{p}_{i}+\boldsymbol{S}_{i}^{(k)}\right\Vert _{2}^{2}\\
+\lambda_{1}\left(\left\Vert \nabla_{H}\boldsymbol{x}\right\Vert _{1}+\left\Vert \nabla_{V}\boldsymbol{x}\right\Vert _{1}\right)
\end{array}\right\} \label{eq:(10)}
\end{equation}

Sub-problem 2: 
\begin{equation}
\boldsymbol{S}^{(k+1)}=\underset{\boldsymbol{S}}{\text{argmin}}\left\{ \begin{array}{c}
\cfrac{1}{2}\stackrel[i=1]{U}{\sum}\left\Vert A_{i}\boldsymbol{x}^{(k+1)}-\boldsymbol{p}_{i}+\boldsymbol{S}_{i}\right\Vert _{2}^{2}\\
+\lambda_{2}\left\Vert \nabla_{V}\boldsymbol{S}\right\Vert _{1}+\lambda_{3}\left\Vert \boldsymbol{S}\right\Vert _{2,1}
\end{array}\right\} \label{eq:(11)}
\end{equation}

For sub-problem 1, the similar alternating minimization method is
used to obtain a solution by solving the following two sub-problems: 
\begin{equation}
\boldsymbol{x}^{(k+\frac{1}{2})}=\underset{\boldsymbol{x}}{\text{argmin}}\left\{ \begin{array}{c}
\cfrac{1}{2}\stackrel[i=1]{U}{\sum}\left\Vert A_{i}\boldsymbol{x}-\boldsymbol{p}_{i}+\boldsymbol{S}_{i}^{(k)}\right\Vert _{2}^{2}\\
+\left\Vert \boldsymbol{x}-\boldsymbol{x}^{(k)}\right\Vert _{2}^{2}
\end{array}\right\} \label{eq:(12)}
\end{equation}
\begin{equation}
\boldsymbol{x}^{(k+1)}=\underset{\boldsymbol{x}}{\text{argmin}}\left\{ \begin{array}{c}
\lambda_{1}\left(\left\Vert \nabla_{H}\boldsymbol{x}\right\Vert _{1}+\left\Vert \nabla_{V}\boldsymbol{x}\right\Vert _{1}\right)\\
+\left\Vert \boldsymbol{x}-\boldsymbol{x}^{(k+\frac{1}{2})}\right\Vert _{2}^{2}
\end{array}\right\} \label{eq:(13)}
\end{equation}
where (\ref{eq:(12)}) can be solved approximately by executing once
SART with the initial value $\boldsymbol{x}^{(k)}$ and projection
data $\boldsymbol{p}_{i}-\boldsymbol{S}_{i}^{(k)}$, and (\ref{eq:(13)})
can be solved by the alternating direction method of multipliers method (ADMM).

For sub-problem 2, the ADMM method is still used.
Firstly, the unconstrained optimization problem (\ref{eq:(11)}) is
converted into the following constrained optimization problem by introducing
two auxiliary variables $\boldsymbol{H}=\nabla_{V}\boldsymbol{S}$ and
$\boldsymbol{W}=\boldsymbol{S}$: 
\begin{equation}
\boldsymbol{S}^{(k+1)}=\underset{\boldsymbol{S}}{\text{argmin}}\left\{ \begin{array}{c}
\cfrac{1}{2}\stackrel[i=1]{U}{\sum}\left\Vert A_{i}\boldsymbol{x}^{(k+1)}-\boldsymbol{p}_{i}+\boldsymbol{S}_{i}\right\Vert _{2}^{2}\\
+\lambda_{2}\left\Vert \nabla_{V}\boldsymbol{S}\right\Vert _{1}+\lambda_{3}\left\Vert \boldsymbol{S}\right\Vert _{2,1}\\
\text{s.t.} \boldsymbol{H}=\nabla_{V}\boldsymbol{S},\boldsymbol{W}=\boldsymbol{S}
\end{array}\right\} \label{eq:(14)}
\end{equation}
Then we can construct its augmented Lagrangian function: 
\begin{align}
 & L(\boldsymbol{S},\boldsymbol{W},\boldsymbol{H},\boldsymbol{\gamma}_{1},\boldsymbol{\gamma}_{2},\mu_{1},\mu_{2})\nonumber \\
 & =\cfrac{1}{2}\stackrel[i=1]{U}{\sum}\left\Vert A_{i}\boldsymbol{x}^{(k+1)}-\boldsymbol{p}_{i}+\boldsymbol{S}_{i}\right\Vert _{2}^{2}\nonumber \\
 & +\lambda_{2}\left\Vert \boldsymbol{H}\right\Vert _{1}+\boldsymbol{\gamma}_{1}^{\top}(\boldsymbol{H}-\nabla_{V}\boldsymbol{S})\nonumber \\
 & +\frac{\mu_{1}}{2}\left\Vert \boldsymbol{H}-\nabla_{V}\boldsymbol{S}\right\Vert _{2}^{2}+\lambda_{3}\left\Vert \boldsymbol{W}\right\Vert _{2,1}\nonumber \\
 & +\boldsymbol{\gamma}_{2}^{\top}(\boldsymbol{W}-\boldsymbol{S})+\frac{\mu_{2}}{2}\left\Vert \boldsymbol{W}-\boldsymbol{S}\right\Vert _{2}^{2}\label{eq:(15)}
\end{align}
where $\mu_{1},\mu_{2}>0$ are the penalty parameters. $\boldsymbol{\gamma}_{1},\boldsymbol{\gamma}_{2}\in  \mathbb{R}^{V\times U}$
are the Lagrangian multipliers. The solution to problem (\ref{eq:(15)})
can be obtained by alternately solving the following five sub-problems:
\begin{equation}
\boldsymbol{S}^{J+1}=\underset{\boldsymbol{S}}{\text{argmin}}\left\{ \begin{array}{c}
\cfrac{1}{2}\stackrel[i=1]{U}{\sum}\left\Vert A_{i}\boldsymbol{x}^{(k+1)}-\boldsymbol{p}_{i}+\boldsymbol{S}_{i}\right\Vert _{2}^{2}\\
+(\boldsymbol{\gamma}_{1}^{\top})^{J}(\boldsymbol{H}^{J}-\nabla_{V}\boldsymbol{S})\\
+\frac{\mu_{1}}{2}\left\Vert \boldsymbol{H}^{J}-\nabla_{V}\boldsymbol{S}\right\Vert _{2}^{2}\\
+(\boldsymbol{\gamma}_{2}^{\top})^{J}(\boldsymbol{W}^{J}-\boldsymbol{S})\\
+\frac{\mu_{2}}{2}\left\Vert \boldsymbol{W}^{J}-\boldsymbol{S}\right\Vert _{2}^{2}
\end{array}\right\} \label{eq:(16)}
\end{equation}
\begin{equation}
\boldsymbol{H}^{J+1}=\underset{\boldsymbol{H}}{\text{argmin}}\left\{ \begin{array}{c}
\lambda_{2}\left\Vert \boldsymbol{H}\right\Vert _{1}\\
+(\boldsymbol{\gamma}_{1}^{\top})^{J}(\boldsymbol{H}-\nabla_{V}\boldsymbol{S}^{J+1})\\
+\frac{\mu_{1}}{2}\left\Vert \boldsymbol{H}-\nabla_{V}\boldsymbol{S}^{J+1}\right\Vert _{2}^{2}
\end{array}\right\} \label{eq:(17)}
\end{equation}
\begin{equation}
\boldsymbol{W}^{J+1}=\underset{\boldsymbol{W}}{\text{argmin}}\left\{ \begin{array}{c}
\lambda_{3}\left\Vert \boldsymbol{W}\right\Vert _{2,1}\\
+(\boldsymbol{\gamma}_{2}^{\top})^{J}(\boldsymbol{W}-\boldsymbol{S}^{J+1})\\
+\frac{\mu_{2}}{2}\left\Vert \boldsymbol{W}-\boldsymbol{S}^{J+1}\right\Vert _{2}^{2}
\end{array}\right\} \label{eq:(18)}
\end{equation}
\begin{equation}
\boldsymbol{\gamma}_{1}^{J+1}=\boldsymbol{\gamma}_{1}^{J}+\mu_{1}(\boldsymbol{H}^{J+1}-\nabla_{V}\boldsymbol{S}^{J+1})\label{eq:(19)}
\end{equation}
\begin{equation}
\boldsymbol{\gamma}_{2}^{J+1}=\boldsymbol{\gamma}_{2}^{J}+\mu_{2}(\boldsymbol{W}^{J+1}-\boldsymbol{S}^{J+1})\label{eq:(20)}
\end{equation}
where superscript $J$ is the number of internal iterations. We omit the superscript $k$ which represents the number of external iterations. The complete algorithm is shown in Algorithm 1.

\makeatletter
\providecommand{\tabularnewline}{\\}
\floatstyle{ruled}
\newfloat{algorithm}{tbp}{loa}
\providecommand{\algorithmname}{Algorithm}
\floatname{algorithm}{\protect\algorithmname}

\begin{algorithm}
Input: real sinogram $\boldsymbol{p}$.

1: Initialize $\boldsymbol{x}^{(0)}$; $\boldsymbol{S}^{(0)}$; iteration
number $K$; regularization parameters $\lambda_{1}$, $\lambda_{2}$, $\lambda_{3}$; $k=0$.

2: while $k<K$:

3: \quad{}\hphantom{}Updating $\boldsymbol{x}^{(k+1)}$ by (\ref{eq:(12)})
and (\ref{eq:(13)}).

4: \quad{}Updating $\boldsymbol{S}^{(k+1)}$ by (\ref{eq:(16)})-(\ref{eq:(20)}).

5: \quad{}$k=k+1$

6: end while

Output: $\boldsymbol{x}$.

\caption{The proposed algorithm.}
\end{algorithm}

\subsection{Experiments}

\begin{figure}[tbh]
\begin{centering}
\subfloat[]{\begin{centering}
\includegraphics[width=0.23\textwidth]{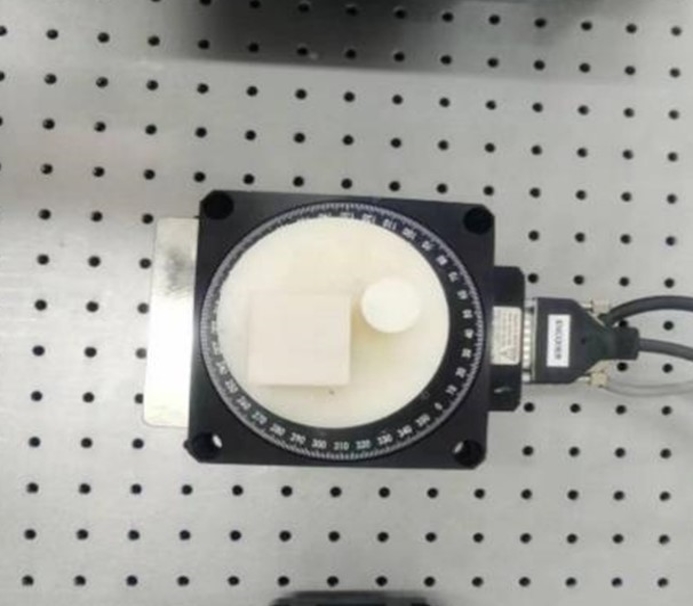}
\par\end{centering}
}\subfloat[]{\begin{centering}
\includegraphics[width=0.23\textwidth]{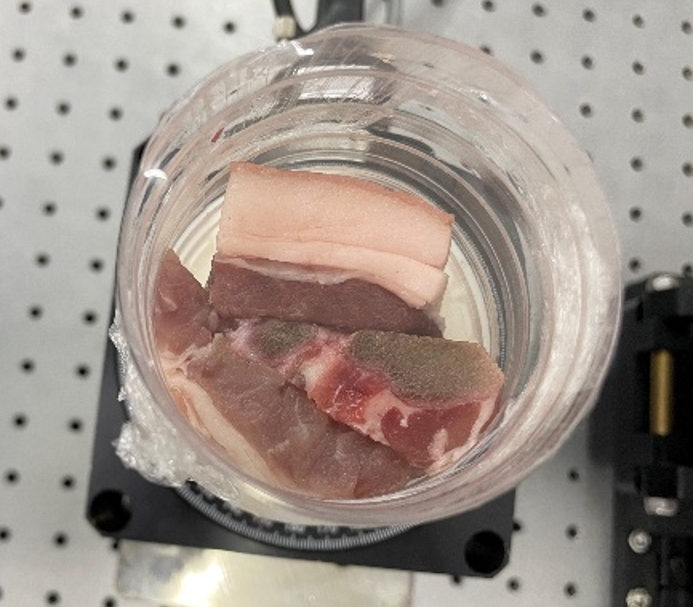}
\par\end{centering}
}
\par\end{centering}
\caption{\label{fig:phantoms}The photograph of two scanned samples. (a) The
water-and-bone sample. (b) The pork-with-bone sample.}
\end{figure}

To evaluate the effectiveness of the proposed method, we perform experiments using two samples, one of which is composed of water and bone equivalent materials (referred to as the water-and-bone sample), and the other is a small piece of pork with bones (referred to as the pork-with-bone sample). The photographs of the two samples are shown in Fig. \ref{fig:phantoms}. It is important to note the particular challenge involved in correcting ring artifacts in the reconstructed image of the water-and-bone sample. The inherent structural simplicity of this sample amplifies the difficulty of the task, making it an ideal candidate to demonstrate the robustness of the proposed correction method. 

\begin{figure}[htbp]
\begin{centering}
\includegraphics[width=0.42\textwidth]{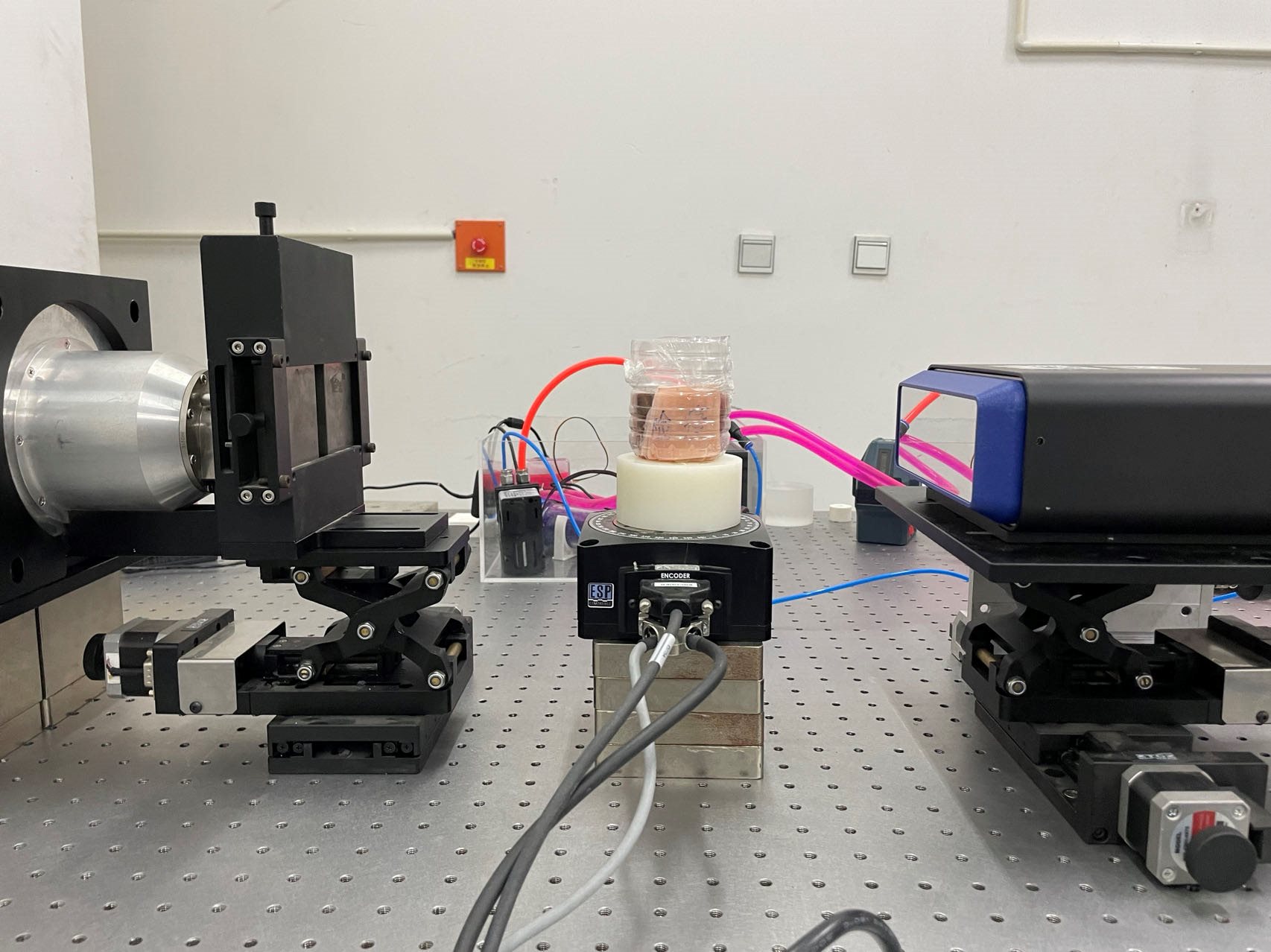}
\par\end{centering}
\caption{\label{fig:system}\textcolor{black}{The experimental CT platform.}}
\end{figure}

The real data is acquired with an experimental platform developed by our laboratory (Fig. \ref{fig:system}), which is equipped with a HAMAMATSU-L10101 X-ray source and an EIGER2 1M-W R-DECTRIS PCD detector. The energy thresholds of the PCD tend to vary among detector units,
resulting in some of the photons being incorrectly counted in different energy bins, which causes particularly severe ring artifacts. This is the reason that we chose a PCD detector to perform the experiments. The scanning parameters are listed in \textcolor{black}{Table \ref{tab:2}}.

For the experiment, only the projection data corresponding to the
central slice is extracted to simulate a fan-beam CT scan. It is noteworthy
that this work can be easily extended to a 3D cone-beam CT scan. Due
to constraints on the length of the current paper, a detailed exploration
of this extension is presented in a separate publication by our research
group.

To demonstrate the benefits of the proposed method, we compare it
with four state-of-the-art methods: the filter-based method proposed
by Münch et al. \cite{munch2009stripe}, the mean projection-based
method proposed by Podgorsak et al. \cite{podgorsak2018use} and Wu
et al. \cite{an2020ring}, and the iterative optimization method proposed
by Salehjahromi et al. \cite{salehjahromi2019new}. For convenience,
we will refer to them as the WF method, Podgorsak method, Wu method,
and Salehjahromi method, respectively. We carefully adjusted their
parameters to ensure the optimal performance of these comparative
methods.

\renewcommand\arraystretch{1.2}
\begin{table}[tbh]
\caption{\label{tab:2}Geometrical parameters of the real scanning device.}
\fontsize{9.6pt}{\baselineskip}\selectfont
\centering{}%
\begin{tabular}{rl}
\toprule 
Parameter  & Value\tabularnewline
\midrule 
\addlinespace
Voltage  & 100 kV\tabularnewline
Current  & 0.3 mA\tabularnewline
Number of views  & 720\tabularnewline
Energy bin thresholds  & 50, 60 keV\tabularnewline
Width of detector unit  & 0.075 mm\tabularnewline
Number of detector units  & 2068$\times$512\tabularnewline
Size of reconstruction image  & 2068$\times$2068\tabularnewline
Distance of X-ray source to detector  & 433.507 mm\tabularnewline
Distance of X-ray source to rotation center  & 205 mm\tabularnewline
\bottomrule
\end{tabular}
\end{table}

\begin{figure*}[htbp]
\begin{centering}
\subfloat[]{\begin{centering}
\includegraphics[width=0.23\paperwidth]{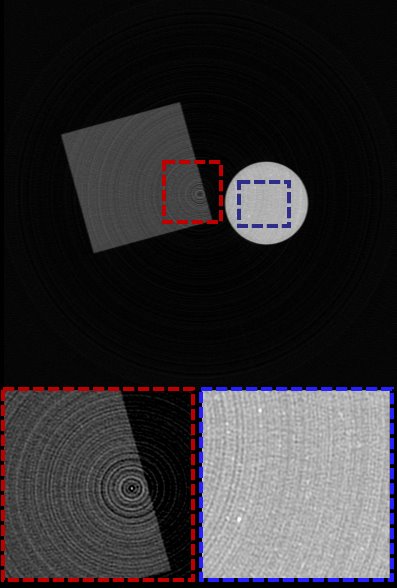}
\par\end{centering}
}\subfloat[]{\begin{centering}
\includegraphics[width=0.23\paperwidth]{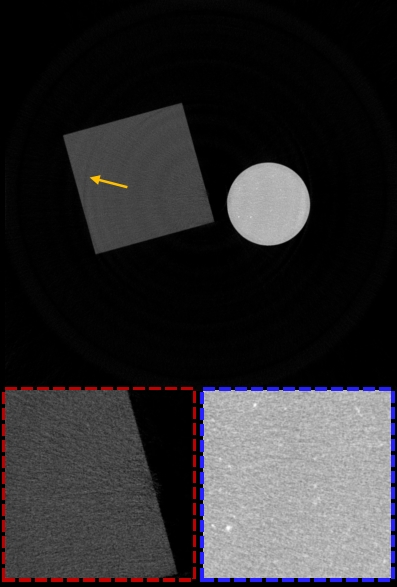}
\par\end{centering}
}\subfloat[]{\begin{centering}
\includegraphics[width=0.23\paperwidth]{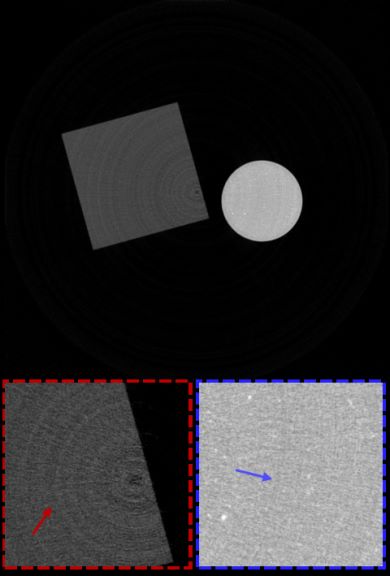}
\par\end{centering}
}
\par\end{centering}
\begin{centering}
\subfloat[]{\begin{centering}
\includegraphics[width=0.23\paperwidth]{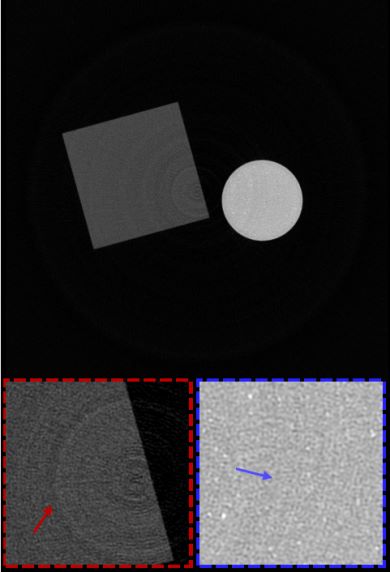}
\par\end{centering}
}\subfloat[]{\begin{centering}
\includegraphics[width=0.23\paperwidth]{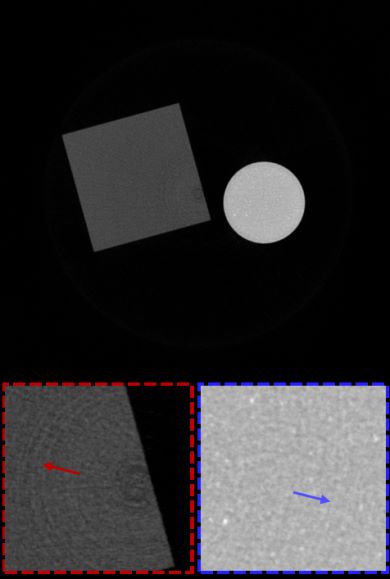}
\par\end{centering}
}\subfloat[]{\begin{centering}
\includegraphics[width=0.23\paperwidth]{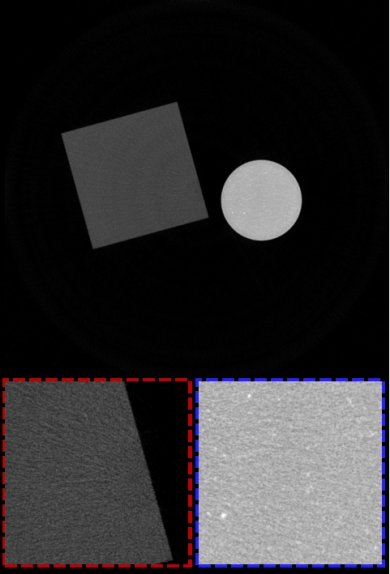}
\par\end{centering}
}
\par\end{centering}
\caption{\label{fig:water-result}The \textcolor{black}{reconstruction} results
of the water-and-bone sample. (a) Uncorrected image; (b) WF method;
(c) Podgorsak method; (d) Wu method; (e) Salehjahromi method; (f)
The proposed method. The display window is set to {[}0, 0.1{]}.}
\end{figure*}

\begin{figure*}[ht]
\begin{centering}
\subfloat[]{\begin{centering}
\includegraphics[width=0.23\paperwidth]{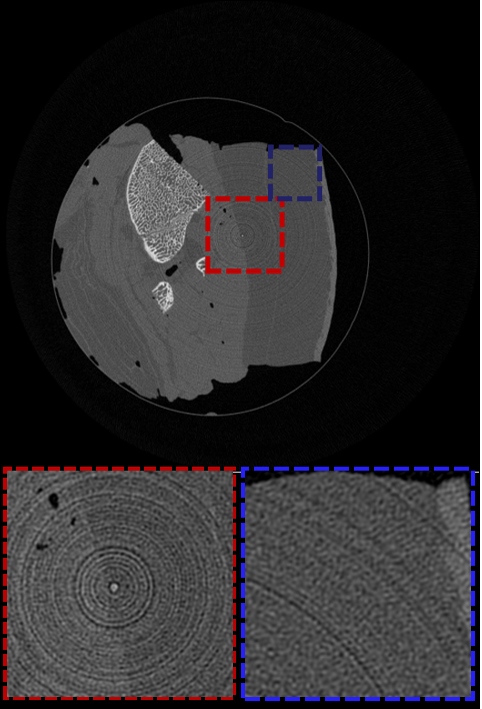}
\par\end{centering}
}\subfloat[]{\begin{centering}
\includegraphics[width=0.23\paperwidth]{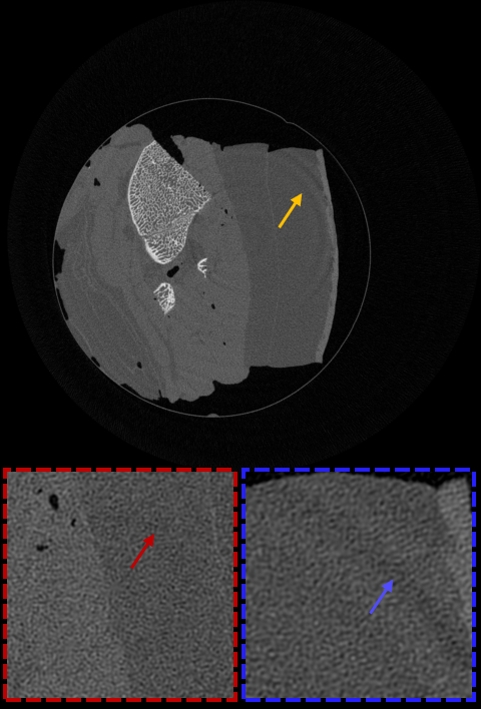}
\par\end{centering}
}\subfloat[]{\begin{centering}
\includegraphics[width=0.23\paperwidth]{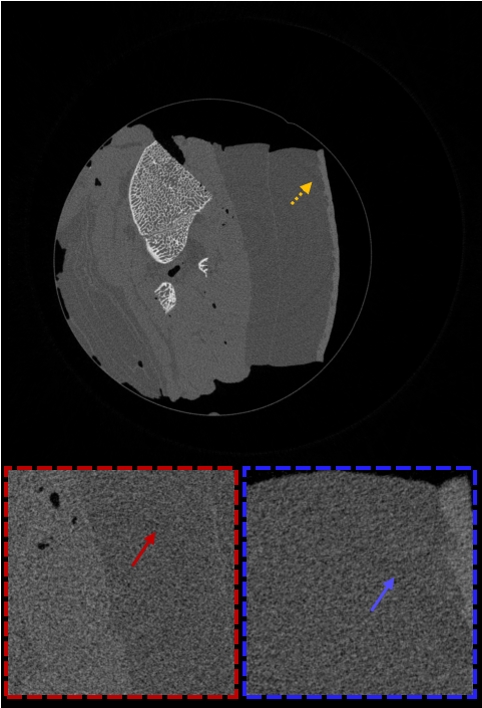}
\par\end{centering}
}
\par\end{centering}
\begin{centering}
\subfloat[]{\begin{centering}
\includegraphics[width=0.23\paperwidth]{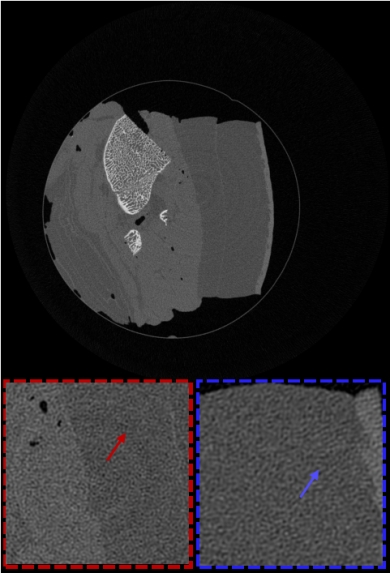}
\par\end{centering}
}\subfloat[]{\begin{centering}
\includegraphics[width=0.23\paperwidth]{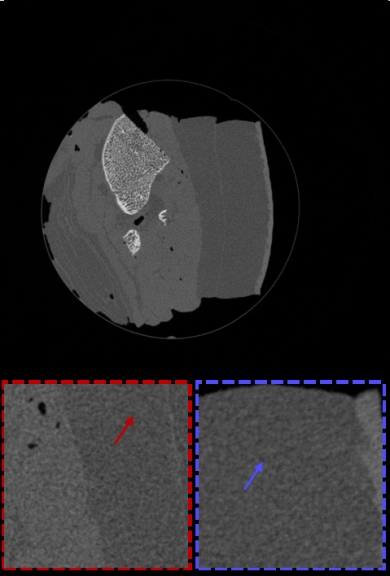}
\par\end{centering}
}\subfloat[]{\begin{centering}
\includegraphics[width=0.23\paperwidth]{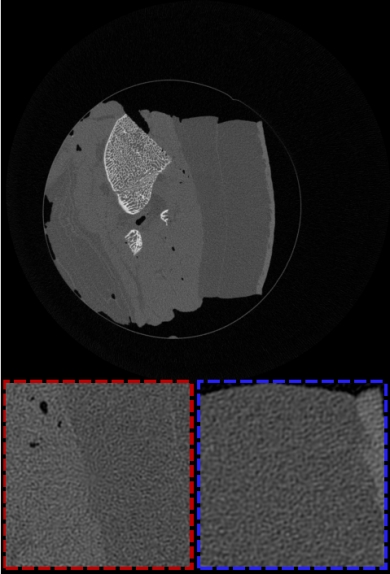}
\par\end{centering}
}
\par\end{centering}
\caption{\label{fig:pork-result} The \textcolor{black}{reconstruction} results
of the pork-with-bone sample. (a) Uncorrected image; (b) WF method;
(c) Podgorsak method; (d) Wu method; (e) Salehjahromi method; (f)
The proposed method. The display window is set to {[}0, 0.08{]}.}
\end{figure*}

\section{Results}

Fig. \ref{fig:water-result} and Fig. \ref{fig:pork-result} show the reconstruction
results of the two samples, including the uncorrected image for comparison
(Fig. \ref{fig:water-result}(a) and Fig. \ref{fig:pork-result}(a)), the
corrected result of the WF method (Fig. \ref{fig:water-result}(b)
and Fig. \ref{fig:pork-result}(b)), the Podgorsak method (Fig. \ref{fig:water-result}(c)
and Fig. \ref{fig:pork-result}(c)), the Wu method (Fig. \ref{fig:water-result}(d)
and Fig. \ref{fig:pork-result}(d)), the Salehjahromi method (Fig. \ref{fig:water-result}(e)
and Fig. \ref{fig:pork-result}(e)), and the proposed method (Fig. \ref{fig:water-result}(f)
and Fig. \ref{fig:pork-result}(f)). Magnified images of regions of interest
(ROIs) are shown at the bottom of each image. The locations of the
ROIs are labeled with blue and red rectangles in Fig. \ref{fig:water-result}(a)
and Fig. \ref{fig:pork-result}(a), respectively.

\subsection{The results of the water-and-bone sample}

Fig. \ref{fig:water-result} shows the reconstructed CT images of
the water-and-bone sample after ring artifact correction. As can be
seen from Fig. \ref{fig:water-result}(a), there are severe ring artifacts
on the reconstructed CT image, which significantly interfere with
the interpretation of the image. The WF method successfully removes
most of the ring artifacts, but it introduces new low-frequency artifacts
(as indicated by the yellow arrow in Fig. \ref{fig:water-result}(b)).
Both the Podgorsak method and the Wu method generally correct the
majority of ring artifacts while preserving image detail, however,
they fall short in addressing wider ring artifacts (as indicated by
the red arrows in Fig. \ref{fig:water-result}(c) and Fig. \ref{fig:water-result}(d)). The Salehjahromi
method effectively corrects the wider ring artifacts, yet some residual
artifacts remain. Furthermore, these methods struggle with handling
ring artifacts that traverse different positions of the scanning sample,
which results in even more pronounced ring artifacts at these locations
(as indicated by the blue arrows). This phenomenon primarily stems from
their lack of consideration for the correlation between the response
inconsistency of detector units and the projection views. In contrast,
the proposed method performs superior in removing nearly all ring
artifacts, particularly excelling in the correction of wider ring
artifacts and those traversing different positions of the sample. 

\subsection{The results of the pork-with-bone sample}

Fig. \ref{fig:pork-result} shows the reconstructed CT images of the
pork-with-bone sample after ring artifact correction. For the correction
results of the WF method, the Eldib method, and the Wu method, most
of the ring artifacts are well removed. However, the WF method introduces
new low-frequency artifacts (as indicated by the yellow arrow in Fig.
\ref{fig:pork-result}(b)). Both the Podgorsak method and the Wu method
have difficulty in correcting those wider ring artifacts (as indicated
by the red arrows in Fig. \ref{fig:pork-result}(c) and Fig. \ref{fig:pork-result}(d)), as well as those ring artifacts that traverse different positions of the sample
simultaneously (as indicated by the blue arrows in Fig. \ref{fig:pork-result}(c)
and Fig. \ref{fig:pork-result}(d)). The Salehjahromi method can effectively correct those wider ring artifacts, but it also fails to correct those ring artifacts
that traverse different positions of the sample simultaneously (as
indicated by the blue arrow in Fig. \ref{fig:pork-result}(e)). As
for the results of the proposed method, there are almost no visible
ring artifacts remaining.

\section{Discussion}
\begin{figure}[htbp]
\begin{centering}
\subfloat[]{\begin{centering}
\includegraphics[width=0.22\textwidth]{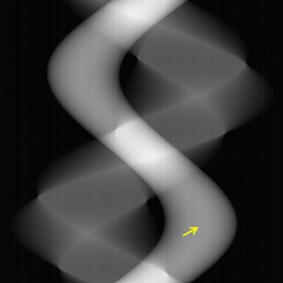}
\par\end{centering}
}\subfloat[]{\begin{centering}
\includegraphics[width=0.22\textwidth]{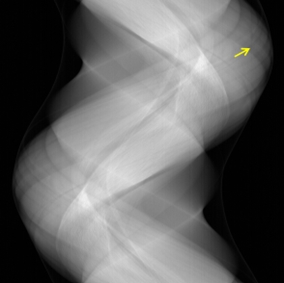}
\par\end{centering}
}
\par\end{centering}
\caption{\label{fig:sinogram}The uncorrected sinogram of two samples. (a) The
uncorrected sinogram of the water-and-bone sample. (b) The uncorrected
sinogram of the pork-with-bone sample.}
\end{figure}

In this paper, we propose a novel dual-domain regularization ring
artifact removal model, which is characterized by considering the
correlation between the response inconsistency of detector units and the projection views, and thus can better correct the ring artifacts, especially the vertical stripe artifacts whose intensity varies greatly in the direction of the projection view. The blue arrows in the ROIs shown in Fig. \ref{fig:water-result} and Fig. \ref{fig:pork-result} highlight the shortcomings of several comparative methods in addressing ring artifacts that traverse different positions of the scanning sample. As can be seen from Fig. \ref{fig:sinogram} that these ring artifacts correspond to the vertical stripe artifacts marked by the yellow arrows in the sinogram, which are characterized by sharp variations along the projection view direction. Since the response inconsistency of the detector unit is related to various factors including the scanning sample and the energy spectrum. Consequently, the response of the same detector unit to X-rays is different when the X-rays pass through different positions of the scanning sample at different projection views. Several comparative methods overlook the correlation between the response inconsistency of the detector units and the projection views, rendering them less effective in correcting such artifacts. Unlike these methods, the proposed method can achieve superior artifact mitigation by considering this correlation.

In some cases, the ring artifacts may be caused by the inconsistency of one whole module. That is to say, the form of ring artifacts is a wide circle, which is challenging to remove completely. The proposed method can still work in these cases. As indicated by the red arrows in the ROIs in Fig. \ref{fig:water-result} and Fig. \ref{fig:pork-result}, while the WF method can mitigate such ring artifacts,
it unfortunately introduces new artifacts that compromise the integrity of image details. The Podgorsak method and the Wu method struggle to remove the particularly wider ring artifacts, likely due to their reliance on filtering methods that inadvertently retain such artifacts as legitimate structural components of the image. It can be seen that both the Salehjahrom method and the proposed method successfully eliminate these wider ring artifacts. We attribute this improvement to the synergistic application of the constraints in both the projection and image domains. Recognizing and rectifying such wide ring artifacts is challenging
when constrained to the projection domain alone; however, employing
the prior information from the reconstructed CT image allows for further correction in the image domain. Therefore, the dual-domain method facilitates a more comprehensive and effective removal of such artifacts.

\section{Conclusion}
\textcolor{black}{An innovative dual-domain regularization model is proposed for ring artifact removal, which corrects stripe artifacts in the projection domain by updating the response inconsistency compensation coefficients of detector units. Simultaneously, it further rectified the ring artifacts by the imposition of sparse constraints on the reconstructed CT image. The effectiveness of the proposed method is validated through experiments conducted on real data. The results demonstrate that the proposed method can remove both high-frequency and low-frequency ring artifacts effectively, and outperforms the competition method in the removal of ring artifacts. }

\section*{Acknowledgment}
The authors are grateful to the Beijing Higher Institution Engineering Research Center of Testing and Imaging as well as the Beijing Advanced Innovation Center for Imaging Technology for funding this research work.

\appendices{}

\bibliographystyle{IEEEtran}
\bibliography{ring}

\begin{thebibliography}{10}
\providecommand{\url}[1]{#1}
\csname url@samestyle\endcsname
\providecommand{\newblock}{\relax}
\providecommand{\bibinfo}[2]{#2}
\providecommand{\BIBentrySTDinterwordspacing}{\spaceskip=0pt\relax}
\providecommand{\BIBentryALTinterwordstretchfactor}{4}
\providecommand{\BIBentryALTinterwordspacing}{\spaceskip=\fontdimen2\font plus
\BIBentryALTinterwordstretchfactor\fontdimen3\font minus \fontdimen4\font\relax}
\providecommand{\BIBforeignlanguage}[2]{{%
\expandafter\ifx\csname l@#1\endcsname\relax
\typeout{** WARNING: IEEEtran.bst: No hyphenation pattern has been}%
\typeout{** loaded for the language `#1'. Using the pattern for}%
\typeout{** the default language instead.}%
\else
\language=\csname l@#1\endcsname
\fi
#2}}
\providecommand{\BIBdecl}{\relax}
\BIBdecl

\bibitem{taguchi2013vision}
K.~Taguchi and J.~S. Iwanczyk, ``Vision 20/20: single photon counting {X}-ray detectors in medical imaging,'' \emph{Medical Physics}, vol.~40, no.~10, p. 100901, 2013.

\bibitem{willemink2018photon}
M.~J. Willemink, M.~Persson, A.~Pourmorteza, N.~J. Pelc, and D.~Fleischmann, ``Photon-counting {CT}: technical principles and clinical prospects,'' \emph{Radiology}, vol. 289, no.~2, pp. 293--312, 2018.

\bibitem{yu2016evaluation}
Z.~Yu, S.~Leng, S.~M. Jorgensen, Z.~Li, R.~Gutjahr, B.~Chen, A.~F. Halaweish, S.~Kappler, L.~Yu, E.~L. Ritman \emph{et~al.}, ``Evaluation of conventional imaging performance in a research whole-body {CT} system with a photon-counting detector array,'' \emph{Physics in Medicine \& Biology}, vol.~61, no.~4, p. 1572, 2016.

\bibitem{schmidt2017spectral}
T.~G. Schmidt, R.~F. Barber, and E.~Y. Sidky, ``A spectral {CT} method to directly estimate basis material maps from experimental photon-counting data,'' \emph{IEEE Transactions on Medical Imaging}, vol.~36, no.~9, pp. 1808--1819, 2017.

\bibitem{persson2012framework}
M.~Persson and H.~Bornefalk, ``A framework for evaluating threshold variation compensation methods in photon counting spectral {CT},'' \emph{IEEE Transactions on Medical Imaging}, vol.~31, no.~10, pp. 1861--1874, 2012.

\bibitem{davis1997x}
G.~Davis and J.~Elliott, ``X-ray microtomography scanner using time-delay integration for elimination of ring artefacts in the reconstructed image,'' \emph{Nuclear Instruments and Methods in Physics Research Section A: Accelerators, Spectrometers, Detectors and Associated Equipment}, vol. 394, no. 1-2, pp. 157--162, 1997.

\bibitem{zhu2013micro}
Y.~Zhu, M.~Zhao, H.~Li, and P.~Zhang, ``Micro-{CT} artifacts reduction based on detector random shifting and fast data inpainting,'' \emph{Medical Physics}, vol.~40, no.~3, p. 031114, 2013.

\bibitem{liu2023detector}
Y.~Liu, C.~Wei, and Q.~Xu, ``Detector shifting and deep learning based ring artifact correction method for low-dose {CT},'' \emph{Medical Physics}, 2023.

\bibitem{van2015dynamic}
V.~Van~Nieuwenhove, J.~De~Beenhouwer, F.~De~Carlo, L.~Mancini, F.~Marone, and J.~Sijbers, ``Dynamic intensity normalization using eigen flat fields in {X}-ray imaging,'' \emph{Optics Express}, vol.~23, no.~21, pp. 27\,975--27\,989, 2015.

\bibitem{kwan2006improved}
A.~L. Kwan, J.~A. Seibert, and J.~M. Boone, ``An improved method for flat-field correction of flat panel {X}-ray detector,'' \emph{Medical Physics}, vol.~33, no.~2, pp. 391--393, 2006.

\bibitem{herman2009fundamentals}
G.~T. Herman, \emph{Fundamentals of computerized tomography: image reconstruction from projections}.\hskip 1em plus 0.5em minus 0.4em\relax Springer Science \& Business Media, 2009.

\bibitem{lifton2019ring}
J.~Lifton and T.~Liu, ``Ring artefact reduction via multi-point piecewise linear flat field correction for {X}-ray computed tomography,'' \emph{Optics Express}, vol.~27, no.~3, pp. 3217--3228, 2019.

\bibitem{seibert1998flat}
J.~A. Seibert, J.~M. Boone, and K.~K. Lindfors, ``Flat-field correction technique for digital detectors,'' in \emph{Medical Imaging 1998: Physics of Medical Imaging}, vol. 3336.\hskip 1em plus 0.5em minus 0.4em\relax SPIE, 1998, pp. 348--354.

\bibitem{bangsgaard2023low}
K.~O. Bangsgaard, G.~Burca, E.~Ametova, M.~S. Andersen, and J.~S. J{\o}rgensen, ``Low-rank flat-field correction for artifact reduction in spectral computed tomography,'' \emph{Applied Mathematics in Science and Engineering}, vol.~31, no.~1, p. 2176000, 2023.

\bibitem{raven1998numerical}
C.~Raven, ``Numerical removal of ring artifacts in microtomography,'' \emph{Review of Scientific Instruments}, vol.~69, no.~8, pp. 2978--2980, 1998.

\bibitem{munch2009stripe}
B.~M{\"u}nch, P.~Trtik, F.~Marone, and M.~Stampanoni, ``Stripe and ring artifact removal with combined {W}avelet—{F}ourier filtering,'' \emph{Optics Express}, vol.~17, no.~10, pp. 8567--8591, 2009.

\bibitem{an2020ring}
K.~An, J.~Wang, R.~Zhou, F.~Liu, and W.~Wu, ``Ring-artifacts removal for photon-counting {CT},'' \emph{Optics Express}, vol.~28, no.~17, pp. 25\,180--25\,193, 2020.

\bibitem{eldib2017ring}
M.~E. Eldib, M.~Hegazy, Y.~J. Mun, M.~H. Cho, M.~H. Cho, and S.~Y. Lee, ``A ring artifact correction method: Validation by micro-{CT} imaging with flat-panel detectors and a 2{D} photon-counting detector,'' \emph{Sensors}, vol.~17, no.~2, p. 269, 2017.

\bibitem{podgorsak2018use}
A.~R. Podgorsak, S.~S. Nagesh, D.~Bednarek, S.~Rudin, and C.~N. Ionita, ``Use of a {CMOS}-based micro-{CT} system to validate a ring artifact correction algorithm on low-dose image data,'' in \emph{Medical Imaging 2018: Physics of Medical Imaging}, vol. 10573.\hskip 1em plus 0.5em minus 0.4em\relax SPIE, 2018, pp. 903--914.

\bibitem{chen2009ring}
Y.~Chen, G.~Duan, A.~Fujita, K.~Hirooka, and Y.~Ueno, ``Ring artifacts reduction in cone-beam {CT} images based on independent component analysis,'' in \emph{2009 IEEE Instrumentation and Measurement Technology Conference}.\hskip 1em plus 0.5em minus 0.4em\relax IEEE, 2009, pp. 1734--1737.

\bibitem{yan2016variation}
L.~Yan, T.~Wu, S.~Zhong, and Q.~Zhang, ``A variation-based ring artifact correction method with sparse constraint for flat-detector {CT},'' \emph{Physics in Medicine \& Biology}, vol.~61, no.~3, p. 1278, 2016.

\bibitem{liang2017iterative}
X.~Liang, Z.~Zhang, T.~Niu, S.~Yu, S.~Wu, Z.~Li, H.~Zhang, and Y.~Xie, ``Iterative image-domain ring artifact removal in cone-beam {CT},'' \emph{Physics in Medicine \& Biology}, vol.~62, no.~13, p. 5276, 2017.

\bibitem{fang2019comparison}
W.~Fang and L.~Li, ``Comparison of ring artifacts removal by using neural network in different domains,'' in \emph{2019 IEEE Nuclear Science Symposium and Medical Imaging Conference (NSS/MIC)}.\hskip 1em plus 0.5em minus 0.4em\relax IEEE, 2019, pp. 1--3.

\bibitem{yuan2021deep}
L.~Yuan, Q.~Xu, B.~Liu, Z.~Wang, S.~Liu, C.~Wei, and L.~Wei, ``A deep learning-based ring artifact correction method for {X}-ray {CT},'' \emph{Radiation Detection Technology and Methods}, pp. 1--11, 2021.

\bibitem{wang2019removing}
Z.~Wang, J.~Li, and M.~Enoh, ``Removing ring artifacts in {CBCT} images via generative adversarial networks with unidirectional relative total variation loss,'' \emph{Neural Computing and Applications}, vol.~31, pp. 5147--5158, 2019.

\bibitem{fu2023deep}
T.~Fu, Y.~Wang, K.~Zhang, J.~Zhang, S.~Wang, W.~Huang, C.~Yao, C.~Zhou, and Q.~Yuan, ``Deep-learning-based ring artifact correction for tomographic reconstruction,'' \emph{Journal of Synchrotron Radiation}, vol.~30, no.~3, 2023.

\bibitem{salehjahromi2019new}
M.~Salehjahromi, Q.~Wang, Y.~Zhang, L.~A. Gjesteby, D.~Harrison, G.~Wang, P.~M. Edic, and H.~Yu, ``A new iterative algorithm for ring artifact reduction in {CT} using ring total variation,'' \emph{Medical Physics}, vol.~46, no.~11, pp. 4803--4815, 2019.

\bibitem{chen2019hyperspectral}
Y.~Chen, W.~He, N.~Yokoya, and T.-Z. Huang, ``Hyperspectral image restoration using weighted group sparsity-regularized low-rank tensor decomposition,'' \emph{IEEE Transactions on Cybernetics}, vol.~50, no.~8, pp. 3556--3570, 2019.

\bibitem{dabov2007image}
K.~Dabov, A.~Foi, V.~Katkovnik, and K.~Egiazarian, ``Image denoising by sparse 3-{D} transform-domain collaborative filtering,'' \emph{IEEE Transactions on Image Processing}, vol.~16, no.~8, pp. 2080--2095, 2007.

\bibitem{buades2005non}
A.~Buades, B.~Coll, and J.-M. Morel, ``A non-local algorithm for image denoising,'' in \emph{2005 IEEE Computer Society Conference on Computer Vision and Pattern Recognition (CVPR'05)}, vol.~2.\hskip 1em plus 0.5em minus 0.4em\relax IEEE, 2005, pp. 60--65.

\end{thebibliography}

\nocite{an2020ring,liang2017iterative,liu2023detector,munch2009stripe,persson2012framework,podgorsak2018use,salehjahromi2019new,taguchi2013vision,van2015dynamic,zhu2013micro,fang2019comparison,raven1998numerical,yuan2021deep,wang2019removing,fu2023deep,buades2005non,chen2009ring,chen2019hyperspectral,dabov2007image,davis1997x,eldib2017ring,kwan2006improved,willemink2018photon,yan2016variation,yu2016evaluation,schmidt2017spectral,herman2009fundamentals,lifton2019ring,seibert1998flat,bangsgaard2023low} 
\end{document}